\documentclass[aps,prb,twocolumn,showpacs,floatfix,superscriptaddress,amsmath,amssymb]{revtex4-2}

\usepackage{amsfonts}
\usepackage{mathrsfs}
\usepackage{amsmath}
\usepackage{xcolor}
\usepackage{graphicx}
\usepackage{bm}
\usepackage{amssymb}
\usepackage{xspace}
\usepackage{epstopdf}
\usepackage{dcolumn}
\usepackage{longtable}
\usepackage{multirow}
\usepackage{float}
\usepackage{comment}
\usepackage[colorlinks=true, letterpaper=true, pdfstartview=FitV,  linkcolor=blue, citecolor=blue, urlcolor=blue]{hyperref}

\makeatother

\begin{document}

\title{Intrinsic nonlinear valley Nernst effect}

\author{Xue-Jin Zhang}
\affiliation{Institute of Applied Physics and Materials Engineering, Faculty of Science and Technology, University of Macau, Taipa, Macau, China}

\author{Jin Cao}
\email{caojin.phy@gmail.com}
\affiliation{Institute of Applied Physics and Materials Engineering, Faculty of Science and Technology, University of Macau, Taipa, Macau, China}

\author{Lulu Xiong}
\affiliation{Institute of Applied Physics and Materials Engineering, Faculty of Science and Technology, University of Macau, Taipa, Macau, China}

\author{Hui Wang}
\affiliation{Science, Mathematics and Technology Cluster, Singapore University of Technology and Design, Singapore 487372, Singapore}

\author{Shen Lai}
\email{laishen@um.edu.mo}
\affiliation{Institute of Applied Physics and Materials Engineering, Faculty of Science and Technology, University of Macau, Taipa, Macau, China}

\author{Cong Xiao}
\email{congxiao@fudan.edu.cn}
\affiliation{Interdisciplinary Center for Theoretical Physics and Information Sciences (ICTPIS), Fudan University, Shanghai 200433, China}

\author{Shengyuan A. Yang}
\affiliation{Research Laboratory for Quantum Materials, Department of Applied Physics, The Hong Kong Polytechnic University, Kowloon, Hong Kong, China}

\begin{abstract}
We investigate the intrinsic nonlinear valley Nernst effect, which induces a transverse valley current via a second-order thermoelectric response to a longitudinal temperature gradient. The effect arises from the Berry connection polarizability dipole of valley electrons and is permissible in both inversion-symmetric and inversion-asymmetric materials. We demonstrate that the response tensor is connected to the intrinsic nonlinear valley Hall conductivity through a generalized Mott relation, with the two being directly proportional at low temperatures, scaled by the Lorenz number. We elucidate the symmetry constraints governing this effect and develop a theory for its nonlocal measurement, revealing a nonlocal second-harmonic signal with a distinct $\rho^2$ scaling. This signal comprises two scaling terms, with their ratio corresponding to the square of the thermopower normalized by the Lorenz number. Key characteristics are demonstrated using a tilted Dirac model and first-principles calculations on bilayer WTe$_2$. Possible extrinsic contributions and alternative experimental detection methods, e.g., by valley pumping and by nonreciprocal directional dichroism, are discussed. These findings underscore the significance of band quantum geometry on electron dynamics and establish a theoretical foundation for nonlinear valley caloritronics.


\end{abstract}

\maketitle

\section{Introduction}
Valley degree of freedom, which is associated with multiple energy extremal points in a semiconducting band structure, has attracted tremendous interest in the past two decades, owing to its potential in realizing novel information storage and processing technologies~\cite{schaibley2016Valleytronics, vitale2018Valleytronics, mak2018Light}. Efficient control of valley degree of freedom is a central task of this emerging field of valleytronics.
Depending on the specific valley structures, several mechanisms have been proposed, such as those based on
valley Hall effect~\cite{xiao2007ValleyContrasting, xiao2012Coupled}, optical selection~\cite{yao2008Valleydependent,xiao2012Coupled}, valley filtering~\cite{rycerz2007Valley, gunlycke2011Graphene, pan2015Perfect}, applied magnetic field~\cite{cai2013Magnetic}, gate field~\cite{Yu2020Valley}, magnetic proximity effect~\cite{qi2015Giant, zhang2016Large}, and etc. Among them, valley Hall effect is one of the earliest proposals~\cite{xiao2007ValleyContrasting, xiao2012Coupled}. It refers to a transverse valley current driven by an
applied electric field. This effect has been extensively studied in several two-dimensional (2D) materials, such as graphene~\cite{Gorbachev2014Detecting, sui2015Gatetunable,Shimazaki2015Generation}
and transition metal dichalcogenides~\cite{mak2014Valley, lee2016Electrical,Wu2019Intrinsic}. There also exists a thermal counterpart of valley Hall effect, namely, the valley Nernst effect (VNE), in which the transverse valley current is driven by a longitudinal temperature gradient instead of electric field~\cite{yu2015Thermally}. VNE has been detected in 2D WSe$_2$~\cite{dau2019Valley, choi2025Berry}. As linear response effects, both valley Hall effect and VNE share the same band geometric origin, from the valley-contrasted Berry curvature.


In recent years, with the increasing interest in nonlinear transport effects, a new research direction --- the nonlinear valleytronics --- starts to emerge. The nonlinear version of valley Hall effect has been proposed~\cite{Yu2014Nonlinear,Rodin2016Valley,das2024Nonlinear,zhou2024Nonlinear,cao2025Theory}, and very recently been observed by nonlocal measurement in a graphene superlattice structure~\cite{HeObservation}. The nonlinear valley Hall effect differs from its linear counterpart in many aspects.
In fact, a key for the experimental detection of nonlinear valley Hall effect in Ref.~\cite{HeObservation} is the recognition that
the nonlinear response exhibits distinct symmetry characters and scaling laws compared to the linear response~\cite{cao2025Theory}.
Particularly, in the nonlinear regime, the direct and inverse valley Hall processes are not reciprocal, and they involve
very different microscopic mechanisms.


Similar to the linear case, nonlinear VNE should also exist. In Ref.~\cite{Yu2014Nonlinear}, Yu \emph{et al.}  proposed
a nonlinear VNE driven by the off-equilibrium change of distribution function in the second order of longitudinal temperature gradient. This is an extrinsic mechanism, since the resulting valley current density $j^v$ is proportional to the square of
electron scattering time $\tau$, i.e., $j^v\propto \tau^2$. In transport studies, the \emph{intrinsic} mechanism,  which gives responses solely determined by the material's inherent band structure properties, is of special interest.
Such intrinsic response coefficients represent characteristic of each material. They can be quantitatively evaluated
using first-principles methods and serve as benchmarks for experimental studies. For nonlinear VNE, there may also exist a contribution from intrinsic mechanism, leading to a $j^v$ independent of $\tau$. However, so far, several important questions have not been clarified for this intrinsic nonlinear VNE: (i) What is the connection between this effect and nonlinear valley Hall effect? (ii)
What is its characteristic signal to be probed in experiment?
(iii) Can this effect have a sizable value in realistic materials? And what is the guideline to search for its suitable material platforms?


In this work, we address the above questions surrounding the intrinsic nonlinear VNE. We present a formula for its response tensor, and show that it is intimately connected to the intrinsic nonlinear valley Hall effect through a generalized Mott relation. This relation reveals that the response coefficients of these two effects are directly proportional, scaled by the Lorenz number.
The symmetry conditions for intrinsic nonlinear VNE are clarified. We develop a theory for the nonlocal measurement of
nonlinear VNE, and analyze the characteristic scaling behavior of the intrinsic contribution that can be probed in experiment.   Our analysis is showcased in a model example and in the realistic material bilayer WTe$_2$. Using first-principles calculations, we show that the effect in bilayer WTe$_2$ is substantial, suggesting it is easily detectable experimentally. In addition, we point out that the valley current resulted from VNE also carries a polarized magnetic quadrupole moment, which can produce discernible signals manifested in nonreciprocal directional dichroism.
These findings not only establish a theoretical foundation for nonlinear valley caloritronics, but also open a new avenue for exploring the physics of intrinsic nonlinear thermoelectric transport in nonmagnetic materials.


\begin{figure}
    \centering
    \includegraphics[width=0.75\linewidth]{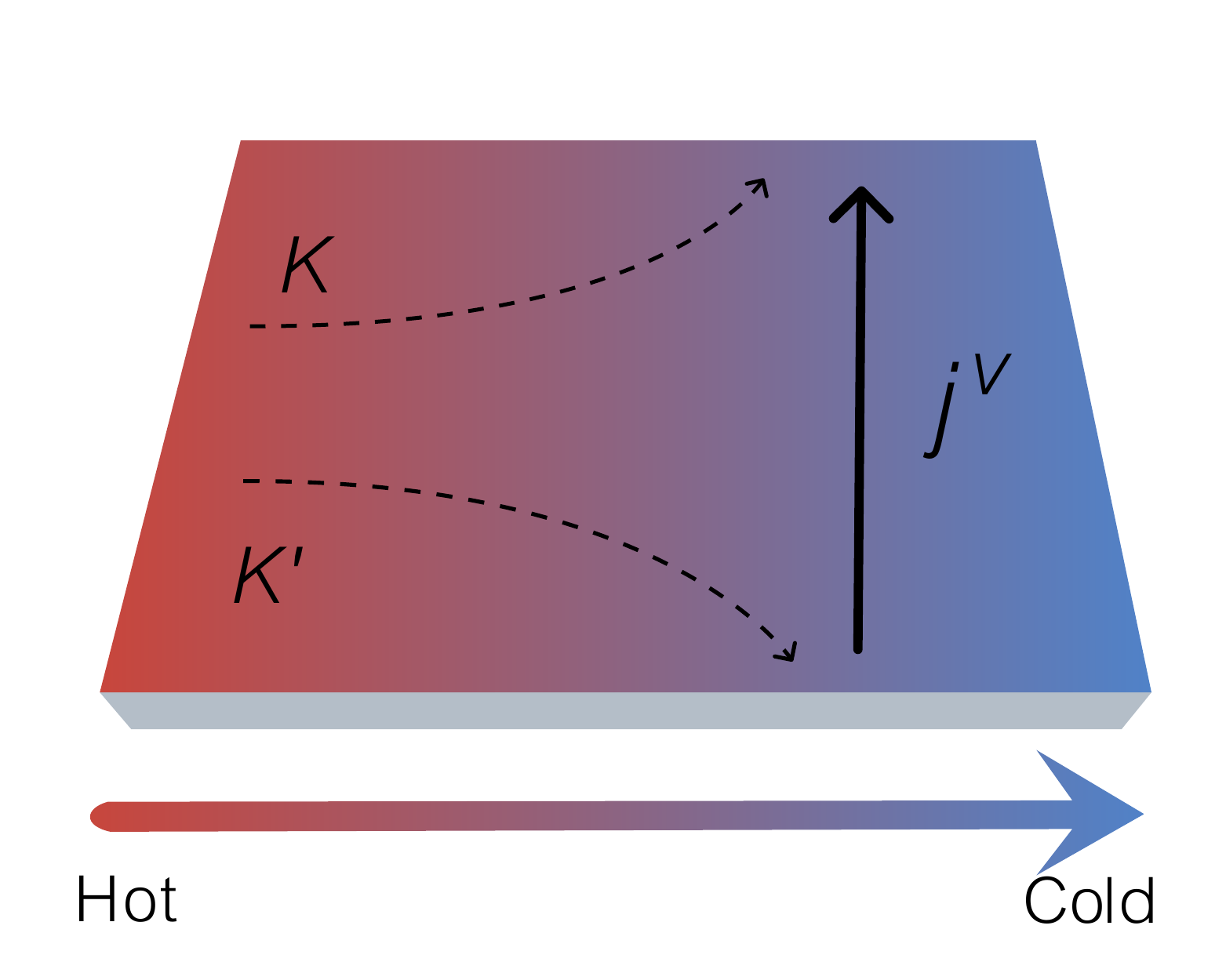}
    \caption{Schematic illustration of VNE. Under a temperature gradient, carriers from two valleys (denoted as $K$ and $K'$) are deflected to opposite transverse directions. For nonlinear VNE, the resulting contribution to Nernst valley current $j^v$ is of second order in the temperature gradient. }
    \label{fig1}
\end{figure}

\section{Theory of intrinsic nonlinear valley Nernst effect}\label{S2}

Consider a 2D nonmagnetic material system with a binary valley degree of freedom connected by time reversal, i.e., there are two energy-degenerate valleys, denoted as $K$ and $K'$, in its low-energy band structure; and the two valleys are switched under time reversal operation $\mathcal T$. The valley current density $j^v$ is defined as
\begin{equation}
  j^v_a=j^K_a-j^{K'}_a,
\end{equation}
where the subscript $a$ (and also $b, c$ below) is a Cartesian index, and $j^{K}$ ($j^{K'}$) is the current density contributed by carriers in $K$ ($K'$) valley.

We are interested in the valley current driven by an applied temperature gradient $\bm\nabla T$. “Nonlinear” in this context means the studied response is of second order in $\bm\nabla T$, so we can express the resulting nonlinear valley current in the following general form
%
%
\begin{equation}
    j_a^{v,\text{nl}} = \beta^{v}_{abc} \partial_b T \partial_c T,
\end{equation}
where $\beta^{v}_{abc}$ is the response tensor and repeated indices are summed over. For Nernst effect, the resulting valley current should be in a direction transverse to the applied temperature gradient. This means the corresponding $\beta^{v}_{abc}$ tensor should be antisymmetric in the first two indices $a$ and $b$ (and by definition, it can always be symmetrized in its last two indices).

The usefulness of valleytronics is based on the condition that the intervalley scattering is weak and the valley index is a well defined quantum number. This condition is also assumed here. The nonlinear valley Nernst conductivity can then be expressed as
\begin{equation}\label{df}
  \beta^{v}_{abc} = \beta^{K}_{abc} - \beta^{K'}_{abc},
\end{equation}
where $\beta^{K}$ is the valley-resolved nonlinear \emph{charge} Nernst conductivity from carriers in valley $K$, similar for  $\beta^{K'}$.

In Ref.~\cite{cao2025Theory}, it was pointed out that in studying valley responses, it is crucial to distinguish valley-even and valley-odd quantities, namely, a quantity is valley-even (-odd) if it is unchanged (changes sign) when the two valleys are switched $K\rightleftharpoons K'$. Clearly, here, the valley current $j^v_a$ and nonlinear VNE conductivity $\beta^{v}_{abc}$ are valley-odd.

Moreover, since the systems considered are assumed to have time reversal symmetry and $\mathcal T$ operation switches the two valleys, the above valley symmetry character will correspond to symmetry properties under  $\mathcal T$ operation.
Specifically, we must have $\beta^{v}_{abc}$ being even under time reversal operation, i.e., it is $\mathcal T$-even.
It follows that only the $\mathcal T$-odd components of $\beta^{K/K'}$, which satisfy
\begin{equation}
    \mathcal{T}: \ \beta^{K}_{abc}\rightleftharpoons -\beta^{K'}_{abc}
\end{equation}
under $\mathcal T$ operation, can contribute to the valley response $\beta^{v}_{abc}$. In contrast, their $\mathcal T$-even components
are canceled out when doing the subtraction in Eq.~(\ref{df}).
For simple notations, here and hereafter, we will just use the symbols $\beta^{K/K'}$ to denote their $\mathcal T$-odd components. Then, we may write
\begin{equation}
    \beta^{v}_{abc}=2\beta^{K}_{abc},
    \label{reduce}
\end{equation}
i.e., to compute the nonlinear valley Nernst response, we just need to find the $\mathcal T$-odd nonlinear charge Nernst response $\beta^K$ of the $K$ valley.

It should be mentioned that the above analysis can be easily extended to inversion symmetry $\mathcal P$. As $\mathcal P$ also switches the two valleys, a similar reasoning shows $\beta^{v}_{abc}$ would be $\mathcal P$-even and only the $\mathcal P$-odd components of $\beta^{K/K'}$ make a contribution. Therefore, the component of charge Nernst response $\beta^K$ that contributes to the valley response $\beta^v$ should be odd under both $\mathcal T$ and $\mathcal P$ operations.

%

The nonlinear charge Nernst conductivity that is both $\mathcal{T}$-odd and $\mathcal{P}$-odd was first studied by Gao and Xiao in Ref.~\cite{gao2018Orbital}. They proposed an intrinsic mechanism for this nonlinear response.
Note that the Nernst effect is driven by a temperature gradient, which necessarily makes the system inhomogeneous. This is very different from the case with electric field driving. The key difference is that in the presence of inhomogeneity, there generally exists a circulating magnetization current $j^\text{mag}$, which does not contribute to transport signals. Therefore, when calculating the transport current $j$, it has to be discounted from the local current density $j^\text{loc}$, i.e., $j_a=j^\text{loc}_a-j^\text{mag}_a$. Up to second order in the gradient of spatial inhomogeneity, one has~\cite{Jackson1999, Groot1972}
\begin{equation}
  \bm j=\bm j^\text{loc}-\bm\nabla\times \bm M+\bm \nabla\times (\partial_a Q_{ab}\hat e_b),
\end{equation}
where $\bm M$ and $Q_{ab}$ are respectively the orbital magnetization vector and the orbital magnetic quadrupole tensor,  and $\hat e_b$ is the unit vector along $b$.
Both $\bm M$ and $Q_{ab}$ can be expressed in terms of band structure properties, and their expressions have been derived in Refs.~\cite{xiao2006BerryPhase,Xiao2020unified,gao2018Orbital}.

Using these results, one finds the following general expression of \emph{intrinsic} transport current, accurate to the second order of statistical force such as $\bm \nabla T$ and $\bm\nabla\mu$ ($\mu$ is chemical potential, and we take $e=\hbar=1$ here)~\footnote{In Ref.~\cite{gao2018Orbital}, the second term of the stated result contains a typo of sign and subscripts.}:
%
%
%
\begin{equation}
    \boldsymbol{j} = - \boldsymbol{\nabla} \times \int [d\boldsymbol{k}] \boldsymbol{\Omega}\mathcal{G}  + \boldsymbol{\nabla} \times \left ( \partial_a \int [d\boldsymbol{k}] \theta_{ab} \mathcal{G} \hat{e}_b \right),\label{jG}
\end{equation}
where
$
  \mathcal{G} = -k_B T \ln[1+e^{- (\varepsilon-\mu)/{k_BT}}]
$
is the grand potential density, $[d\boldsymbol{k}]$ denotes the integration over momentum and the summation over band index,
and for simplicity, the dependence of quantities (under the integrals) on band index $n$ and momentum $\bm k$ is not written out explicitly.
The two terms in Eq.~(\ref{jG}) are respectively the first-order and the second-order responses. The first order response is connected to the Berry curvature $\boldsymbol{\Omega}$ \cite{xiao2006BerryPhase}. The second order response, which is the focus here, is connected to the Berry connection polarizability (BCP) dipole
\begin{equation}
  \theta_{ab} = \epsilon_{bcd}v_{c}G_{ad},
\end{equation}
where the BCP tensor for band $n$ and momentum $\bm k$ is given by~\cite{gao2014Field, liu2022Berry}
\begin{equation}\label{Gab}
    G_{ab}^n(\bm k) = 2\text{Re} \sum_{m\neq n} \frac{\mathcal{A}_a^{nm} \mathcal{A}_b^{mn}}{\varepsilon_n - \varepsilon_m},
\end{equation}
$\mathcal{A}_a^{nm} = i\langle u_n | \partial_{k_a} | u_m \rangle$ is the interband Berry connection, $|u_n \rangle$  is the cell-periodic part of the Bloch state, with band energy $\varepsilon_n$ and band velocity $\boldsymbol{v}^n = \partial \varepsilon_n / \partial \boldsymbol{k}$. Again, it is understood that the quantities on the right hand side of (\ref{Gab}) are taken at the same $\bm k$.
The BCP tensor, like the Berry curvature, is a gauge-invariant geometric property of the electronic band structure~\cite{liu2022Berry}.
Physically, it represents the Berry connection or the positional shift of the electron wave packet induced by the electric field. Previous studies have demonstrated its significance in various effects, such as nonlinear charge transport~\cite{gao2014Field}, nonlinear spin responses~\cite{xiao2022Intrinsic, wang2025Intrinsic}, and nonlinear layer response \cite{zheng2024interlayer}.

Now, according to Eq.~(\ref{reduce}) and our analysis above, to find $\beta^v$ for intrinsic nonlinear valley response, we just need to apply Eq.~(\ref{jG}) to the $K$ valley and retain the terms that are second order in $\bm \nabla T$. A straightforward calculation gives
%
%
%
%
\begin{equation}
    \beta^{v}_{abc} = 2\int_K [d\boldsymbol{k}] \left (v_{a}^{n}G_{bc}^{n}-v_{b}^{n}G_{ac}^{n}\right) \frac{ (\varepsilon_{n}-\mu)^{2}}{T^2}f_0', \label{NVNE}
\end{equation}
where the integral is over states of $K$ valley, $f_0$ is the Fermi-Dirac distribution, and in this formula, we have restored the band index $n$ of the quantities but still kept $\bm k$ dependence implicit. By introducing a valley index $\lambda=\pm 1$ for states in $K/K'$ valley, the above formula can also be cast into the following form
\begin{equation}
    \beta^{v}_{abc} = \int [d\boldsymbol{k}] \lambda\left (v_{a}^{n}G_{bc}^{n}-v_{b}^{n}G_{ac}^{n}\right) \frac{ (\varepsilon_{n}-\mu)^{2}}{T^2}f_0',
\end{equation}
where the integral is over both valleys.

We have a few remarks. First, this nonlinear valley thermoelectric conductivity is clearly intrinsic, i.e., it is independent of scattering and is determined solely by the band structure. Specifically, the formula Eq.~(\ref{NVNE}) shows it is determined by the BCP dipole of $K$ valley weighted by a factor $(\varepsilon_n-\mu)^2/T^2$ on the Fermi surface.
Hence, it represents an inherent property of each valley material platform.
This formula can be readily implemented in first-principles calculations to compute the response coefficients for real materials.

Second, the appearance of $f_0'$ in the formula shows the response is a Fermi surface property, as it should be for a bulk transport process. To have well-defined valley degree of freedom, one should consider a doping range where Fermi surfaces of the two valleys are sufficiently separated in momentum space.

Third, in formula (\ref{NVNE}), to make the expression concise, we did not explicitly symmetrize the last two indices $b$ and $c$ (which can always be done to obtain the symmetric component $\beta^{v}_{a(bc)}$). Meanwhile, it is important to note that $\beta^{v}$ is already  antisymmetric in its first two indices $a$ and $b$ (i.e., it is automatically the nonlinear valley Nernst response we are looking for). This indicates that the intrinsic response must be of Hall type, i.e., the valley current resulted from intrinsic mechanism must be in the direction transverse to the applied temperature gradient.
This has been shown as a general feature in both the linear and nonlinear cases.


\section{Nonlinear Mott relation}

Mott relation establishes a connection between electric conductivity $\sigma$ and thermoelectric response coefficient $\xi$ in linear response. These quantities are second rank tensors defined by
\begin{equation}
  j_a=\sigma_{ab}E_b+\xi_{ab}\partial_b T,
\end{equation}
where $j_a$ is the charge current. Mott relation states that at low temperature with $k_B T\ll \mu$~\cite{marder2010},
\begin{equation}\label{lMott}
  \xi_{ab}=\frac{\pi^2 k_B^2}{3e}T\sigma_{ab}',
\end{equation}
where $\sigma$ is the zero-temperature conductivity and the derivative is with respect to the Fermi energy $\varepsilon_F$ ($= \mu$ at zero temperature). Here, we have temporarily restored the electric charge $e$ $(>0)$ in the relation.

For the nonlinear valley transport considered here, we may also seek a connection between the nonlinear thermoelectric response tensor $\beta^v$ and the nonlinear valley conductivity $\chi^v$, which is defined by
\begin{equation}
  j^{v,\text{nl}}_a=\chi_{abc}^v E_b E_c.
\end{equation}
Again, corresponding to $\beta^v$ in (\ref{NVNE}), we shall focus on $\chi^v$ from the intrinsic mechanism.
This intrinsic $\chi^v$ can be obtained from two different ways.

The first way is analogous to what we have done in Sec.~\ref{S2}
for $\beta^v$, namely, $\chi^v$ can be obtained from the $\mathcal T$-odd nonlinear charge conductivity $\chi^K$ for the $K$ valley:
\begin{equation}
  \chi^v_{abc}=2\chi^K_{abc}.
\end{equation}
The theory for intrinsic nonlinear charge conductivity, which is $\mathcal T$-odd, has been developed in Refs.~\cite{gao2014Field, wang2021Intrinsic, liu2021Intrinsic}. Using the result and applying it to the $K$ valley here, one obtains
\begin{equation}
    \chi_{abc}^v = 2\int_K [d\boldsymbol{k}] \left (v_{a}^{n}G_{bc}^{n}-v_{b}^{n}G_{ac}^{n}\right) f_0'.
    \label{NVHE}
\end{equation}

The second way is to start from the formula (\ref{jG}). We mentioned in the discussion above Eq.~(\ref{jG}) that the formula also applies to the case where the driving force is the chemical potential gradient $\bm\nabla \mu$. Via straightforward calculation, one finds that at second order in this gradient, i.e., for the nonlinear valley current
\begin{equation}
    j_a^{v,\text{nl}} = \tilde\chi^{v}_{abc} \partial_b \mu\, \partial_c \mu,\label{jmu}
\end{equation}
the response tensor is given by
\begin{equation}
    \tilde\chi_{abc}^v = 2\int_K [d\boldsymbol{k}] \left (v_{a}^{n}G_{bc}^{n}-v_{b}^{n}G_{ac}^{n}\right) f_0'.\label{mu}
\end{equation}
One notices that $\chi_{abc}^v$ in (\ref{NVHE}) and $\tilde\chi_{abc}^v$ in (\ref{mu}) have the same expression (they differ by a constant factor $e^2$, if $e$ is restored). This is not a coincidence. According to Einstein relation, $\bm\nabla\mu$ are $\bm E$ are equivalent in driving electric current. Making the replacement $\partial_a\mu\rightarrow E_a$ in (\ref{jmu}), one immediately finds that $\chi^v=\tilde\chi^v$.

It is very nice to achieve a consistency between the two different approaches. This consistency with fundamental physical relations, i.e., the Einstein relation here, further strengthens the validity of the whole theoretical framework.


Now, having $\beta^v$ given in Eq.~(\ref{NVNE}) and $\chi^v$ in Eq.~(\ref{NVHE}), one observes an apparent resemblance between the two. Both involve BCP dipole on the Fermi surface. Hence, it is natural to expect that the two are connected by some relation, similar to the Mott relation connecting $\sigma$ and $\xi$ in linear response.
Indeed, by a manipulation as shown in Appendix~A, we find that
\begin{equation}
        \beta^{v}_{abc}=\int d\varepsilon \Big(-\frac{\partial f}{\partial\varepsilon}\Big)\frac{ (\varepsilon-\mu)^{2}}{T^{2}}\chi^{v}_{abc} (\varepsilon),
        \label{generalized nonlinear Mott}
\end{equation}
%
%
%
where $\chi^v_{abc} (\varepsilon)$ is the zero-temperature nonlinear valley conductivity with Fermi energy at $\varepsilon$:
\begin{equation}
  \chi^v_{abc} (\varepsilon)=-2\int_K\left[d\boldsymbol{k}\right]\delta (\varepsilon-\varepsilon_{n})\left (v^n_{a}G^n_{bc}-v^n_{b}G^n_{ac}\right).
\end{equation}

At low temperatures with $k_BT\ll \mu$, by the standard Sommerfeld expansion, i.e., expanding $\chi^v$ around $\varepsilon_F$ ($=\mu$) and performing the integration over $\varepsilon$, we obtain the following nonlinear Mott relation ($e$ is recovered here):
%
\begin{equation}
    \beta^v_{abc}=\mathcal{L}\chi^v_{abc},
    \label{standard nonlinear Mott}
\end{equation}
where $\mathcal{L}=\frac{\pi^{2}k_{B}^{2}}{3e^{2}}=2.43\times 10^{-8}$ W$\cdot \Omega\cdot$K$^{-2}$ is the Lorenz number.
According to this relation, at low temperatures, the two nonlinear response tensors are not independent, but intimately connected to each other by a physical constant $\mathcal{L}$.

We have a few remarks before proceeding. First, the nonlinear Mott relation in (\ref{standard nonlinear Mott})
has a different structure compared to the linear Mott relation in (\ref{lMott}). In the linear case, the thermoelectric response tensor is related to the derivative of electric conductivity, but in the nonlinear case, the two tensors are proportional to each other, not involving derivatives. Actually, the structure of nonlinear Mott relation is
the same as the Wiedemann-Franz law in the linear case, which connects $\sigma$ to the electronic thermal conductivity $\kappa$, i.e., $\kappa_{ab}=\mathcal{L}\sigma_{ab}$~\cite{marder2010}. Mathematically, this is because $\sigma$ and $\kappa$ also satisfy a relation similar to Eq.~(\ref{generalized nonlinear Mott}).

Second, the linear Mott relation involves the derivative $\sigma'$. In real cases, this derivative might be sensitive to the system details, especially around van Hove singularities or when the variation comes from the energy dependence of scattering time; and it is difficult to determine the value of derivative with precision. Hence, the linear Mott relation (\ref{lMott}) usually does not have a good accuracy and it is hard to use in practice~\cite{marder2010, Xiao2016PRB}. In contrast, the nonlinear Mott relation does not involve a derivative, so like Wiedemann-Franz law. One can expect it is much more useful than its linear counterpart.

Third, this form of nonlinear Mott relation in (\ref{standard nonlinear Mott}) has also been found for the $\mathcal T$-even nonlinear responses arising from extrinsic Berry curvature dipole mechanism~\cite{Zeng2020}. Here, we demonstrate the relation for the intrinsic $\mathcal T$-odd nonlinear responses. In fact, one can expect that the relation should be more general. At least on the level of relaxation time approximation, this can be seen by noting that in the kinetic equation, $\bm\nabla \mu$ has equal status as the combination $(\varepsilon_n-\mu)\bm\nabla T/T$ (and by Einstein relation $\bm\nabla\mu$ can be replaced by $E$ field), which makes the appearance of relation (\ref{generalized nonlinear Mott}) quite natural. The range of validity of this nonlinear Mott relation would be an interesting topic to investigate in future studies.

\begin{table}[t!]
    \centering
    \setlength{\tabcolsep}{6.2pt}
    \renewcommand{\arraystretch}{1.5}
    \begin{tabular}{ccccccccc}
    \hline
    \hline
      & $\mathcal{M}_x$
      & $\mathcal{M}_z$
      & $\mathcal{C}_{2x}$
      & $\mathcal{C}_{3z/4z}$
      & $\mathcal{P^\prime}$
      & $\mathcal{M}_x^{\prime}$
      & $\mathcal{C}_{2x}^{\prime}$
      & $\mathcal{C}_{2z}^{\prime}$  \\
    \hline
    $\mathcal{X}^v_{zy}$
      & $ \times $
      & \checkmark
      & \checkmark
      & $ \times $
      & \checkmark
      & \checkmark
      & $ \times $
      & \checkmark \\
    $\mathcal{X}^v_{zx}$
      & \checkmark
      & \checkmark
      & $ \times $
      & $ \times $
      & \checkmark
      & $ \times $
      & \checkmark
      & \checkmark \\
    \hline
    \hline
    \end{tabular}
    \caption{Symmetry constraint on the intrinsic nonlinear VNE conductivity.
    A check mark (\checkmark) indicates the symmetry allows the response component,
    whereas a cross ($\times$) indicates the component is prohibited. The symmetry symbol without (with) a prime means the operation is of valley-preserve (valley-switch) type (see discussion in the main text).}
    \label{tab:symmetry}
\end{table}

\section{Symmetry analysis}

Under inversion $\mathcal P$, the valley current is unchanged but the temperature gradient reverses its direction. As such, the linear VNE is forbidden in inversion symmetric systems, but the nonlinear VNE may be allowed and becomes the leading order effect.

Besides the symmetries $\mathcal{T}$ and $\mathcal{P}$, the crystalline symmetries will impose constraints on the  response tensor of nonlinear VNE.
As mentioned, for the intrinsic response studied here, the corresponding tensor
$\beta^{v}_{abc}$ is already antisymmetric with respect to the indices $a$ and $b$. Because of this antisymmetry, $\beta^{v}_{abc}$ can be equivalently represented by a simplified second rank tensor
\begin{equation}
    \mathcal{X}^v_{cd} = \epsilon_{abc} \beta^v_{abd}/2,
\end{equation}
where $\epsilon_{abc}$ is the Levi-Civita symbol. And the symmetry analysis can be performed on this $\mathcal{X}^v$ tensor.

It is important to note that $\mathcal{X}^v_{cd}$ is not simply a rank-2 pseudo-tensor. As pointed out in Ref.~\cite{cao2025Theory}, when exploring valley response properties, it is crucial to distinguish whether a symmetry operation interchanges or preserves the valley index. Based on this, all symmetry operations of the system can be put into two classes: valley-switch vs. valley-preserve. Such a distinction is necessary, because for valley-odd quantities, e.g., the response coefficients for VNE (no matter linear or nonlinear), an extra minus sign must be added under valley-switch operations, in addition to the usual transformation of a rank-2 pseudo-tensor.

The symmetry transformation property of $\mathcal{X}^v$ can thus be expressed as follows:
\begin{equation}
    \mathcal{X}^v = \eta_v \text{det} (\mathcal{O}) \mathcal{O} \mathcal{X}^v \mathcal{O}^{-1},
\end{equation}
where $\mathcal{O}$ is a point group operation, and $\eta_v=\pm$ if $\mathcal{O}$ is of valley-preserve/valley-switch type.

The obtained constraints by typical point group symmetry operations on the intrinsic nonlinear valley Nernst conductivity are presented in Table~\ref{tab:symmetry}. For each operation, we explicitly indicate whether it belongs to the valley-preserve or valley-switch type, by adding a prime ($\mathcal{O}^\prime$) for valley-switch operations. It should be noted that whether a specific point group operation is valley-preserve or valley-switch usually depends on the location of valleys in momentum space~\cite{cao2025Theory}. Nevertheless, for two valleys connected by $\mathcal{T}$ operation, $\mathcal{P}$ and $\mathcal{C}_{2z}$ are always of valley-switch type, because their action in momentum space is the same as $\mathcal{T}$. On the other hand, $\mathcal{M}_z$ is always valley-preserve for 2D systems considered here.
The constraints obtained in Table~\ref{tab:symmetry} will be useful for analyzing valley response of specific materials and offer guidance for the design of valleytronic platforms.

\section{application to tilted Dirac model}

To understand the features of intrinsic nonlinear VNE, we apply our theory to a 2D tilted Dirac model.
The model consists of two $\mathcal T$-connected Dirac valleys, labeled by valley index $\lambda=\pm 1$ and assumed to be sufficiently separated in momentum space. The model reads
\begin{equation}
H=\lambda wk_{x}+\lambda v\left(k_{x}\tau_{x}+k_{y}\tau_{y}\sigma_{y}\right)+\Delta\tau_{z},
\end{equation}
where $\boldsymbol{k}$ is measured from the band extreme (i.e., the valley center), $\tau$'s and $\sigma$'s are Pauli matrices, $\Delta$ opens a small band gap, and $w$ introduces a tilt along $k_{x}$ direction. The system as a whole preserves both $\mathcal{T}$ and $\mathcal{P}$ symmetries. The two valleys tilt in opposite directions, and they are related by either $\mathcal{P}=\tau_{0}\sigma_{0}$ or $\mathcal{T}=i\sigma_{y}$.
This also means a \emph{single} valley does not preserve $\mathcal{T}$ or $\mathcal{P}$ symmetry by itself; it only has
the combined $\mathcal{PT}=i\sigma_{y}$ symmetry. In addition, a single valley here also possesses
$\mathcal{M}_{y}=i\sigma_{x}$ and $\mathcal{M}_{z}=i\sigma_{y}$ symmetries.

\begin{figure}
    \centering
    \includegraphics[width=\linewidth]{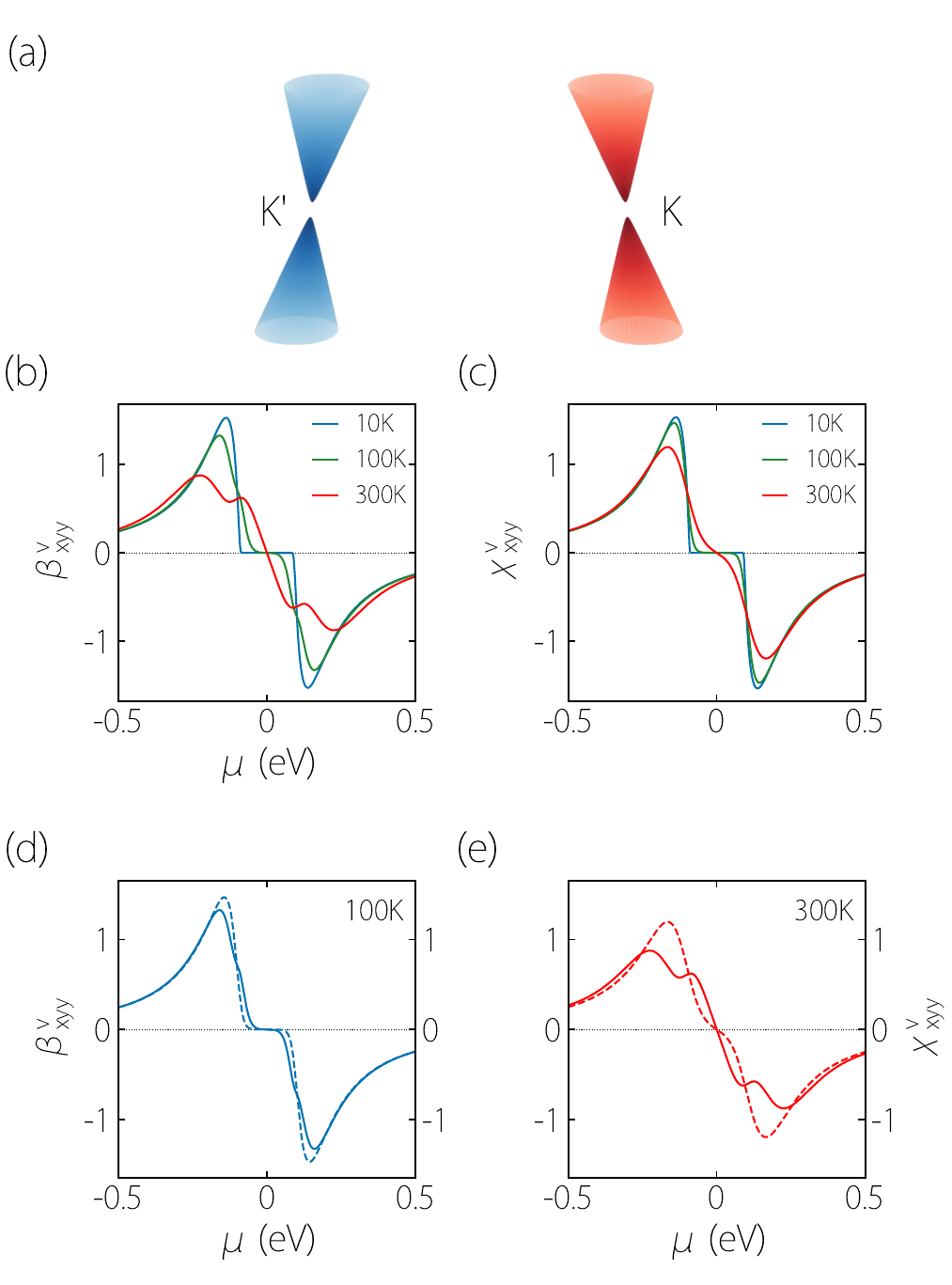}
    \caption{Intrinsic nonlinear VNE and nonlinear Mott relation for the tilted Dirac model. (a) The spectrum of the model, which consists of two $\mathcal{T}$-connected tilted Dirac valleys. (b) Calculated intrinsic nonlinear valley Nernst conductivity $\beta_{xyy}^{v}$ at three different temperatures. (c) Calculated intrinsic nonlinear valley Hall conductivity $\chi_{xyy}^{v}$ at three different temperatures. Here, to check nonlinear Mott relation,  $\beta$ and $\chi$ are put in units of $\pi ek^2_Bw/(96\Delta^2)$ and $e^3w/(32\pi\Delta^2)$, respectively. The two units differ by the Lorenz number, so that $\beta$ and $\chi$ in these figures can be directly compared. (d,e) show such a comparison at (d) low and (e) high temperatures, where the solid (dashed) line is for $\beta_{xyy}^{v}$ ($\chi_{xyy}^{v}$). In the calculation, we take $v=1\times10^6$~m/s, $w=0.4v$, and $\Delta=0.1$~eV.}
    \label{fig_dirac}
\end{figure}

The band dispersion is given by
\begin{equation}
  \varepsilon^{\left(\lambda,\pm\right)}(\bm k)=\lambda wk_{x}\pm\sqrt{v^{2}k^{2}+\Delta^{2}}.
\end{equation}
And each band is doubly degenerate due to the $\mathcal{PT}$ symmetry. This band structure of the model is schematically illustrated in Fig.~\ref{fig_dirac}(a).

The linear VNE is forbidden by $\mathcal{P}$ symmetry in this model. However, the nonlinear VNE is allowed. According to Table~\ref{tab:symmetry}, neither $\mathcal{M}_{z}$ nor $\mathcal{P}$ forbids the nonlinear VNE. Due to valley-preserve symmetry $\mathcal{M}_{y}$, the only nonzero response tensor component for this model is $\mathcal{X}_{zy}^{v}=\beta_{xyy}^{v}$.

The calculated $\beta_{xyy}^{v}$ using Eq.~(\ref{NVNE}) is plotted in Fig.~\ref{fig_dirac}(b) as a function of chemical potential. For comparison, we also plot the intrinsic nonlinear valley Hall conductivity $\chi_{xyy}^{v}$ in Fig.~\ref{fig_dirac}(c).
One has the following observations.
(i) At low temperatures, both $\beta_{xyy}^{v}$ and $\chi_{xyy}^{v}$ are zero within the band gap, indicating a vanishing response in insulating state. (ii) A response peak appears near the band edges, which can be attributed to the enhancement of BCP around small-gap region. (iii) The response changes sign across the band gap. (iv) At elevated temperatures, the single peak at band edge starts to split into two. This can be traced to the factor $(\varepsilon_n-\mu)^2 f_0'/T^2$ in Eq.~(\ref{NVNE}), which exhibits this characteristic change when $T$ increases.

In Figs.~\ref{fig_dirac}(d) and (e), we compare $\beta_{xyy}^{v}$ obtained from Eq.~(\ref{NVNE}) (solid line) and the result from using the nonlinear Mott relation Eq.~(\ref{standard nonlinear Mott}) (dashed line). One can see that the nonlinear Mott relation holds well at low temperature (see Fig.~\ref{fig_dirac}(d)), regardless of the position of chemical potential relative to band edges (which are van Hove singularities).
Meanwhile, at elevated temperatures, the derivation from nonlinear Mott relation starts to appear (see Fig.~\ref{fig_dirac}(e)). Particularly, the double peak feature is not captured by the result from nonlinear Mott relation.

\begin{figure}[b]
    \centering
    \includegraphics[width=\linewidth]{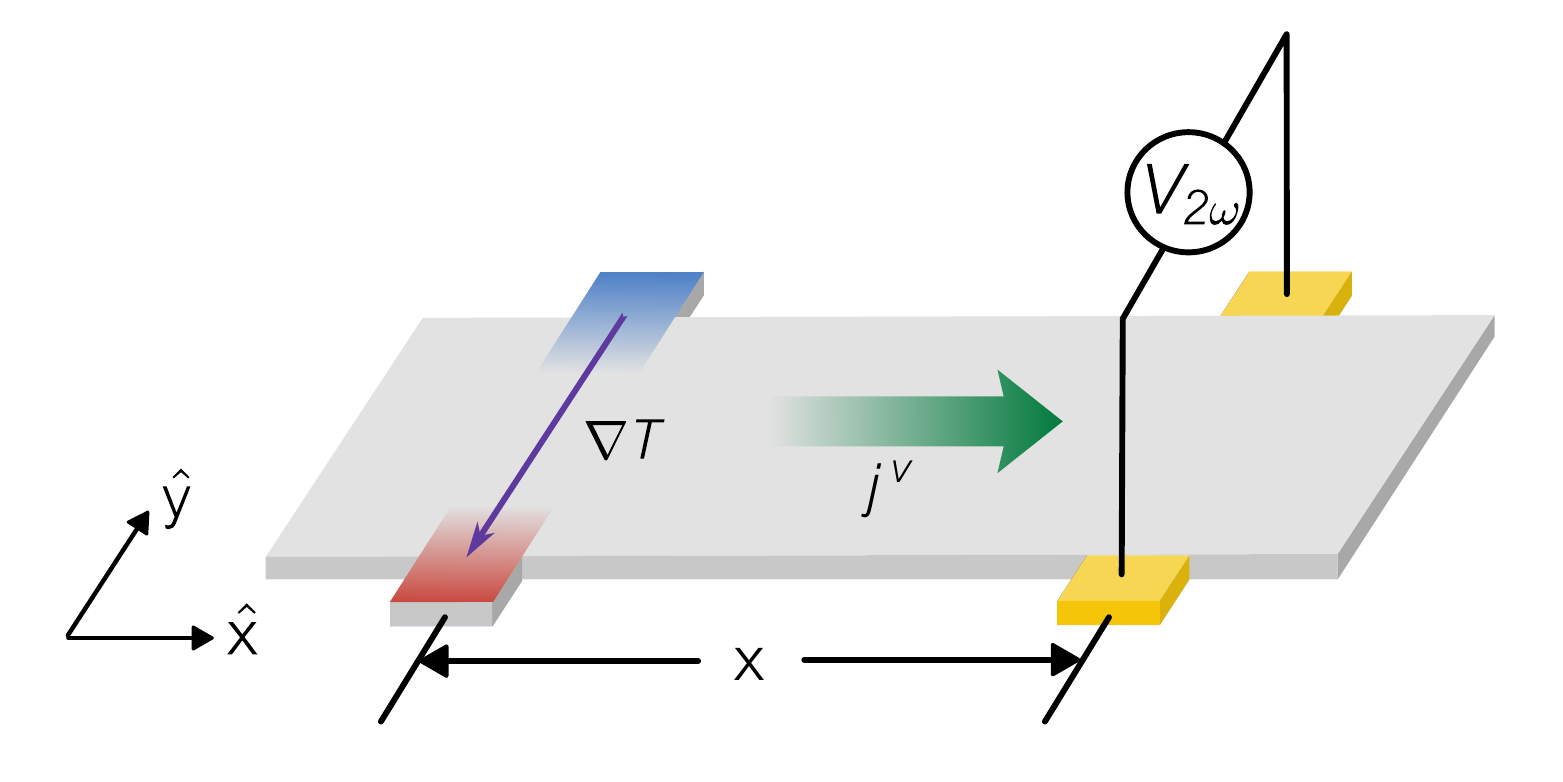}
    \caption{Schematic illustration of a nonlocal measurement setup. See the main text for a description. }
    \label{fig_nonlocal}
\end{figure}

\section{Nonlocal transport signal}

After presenting the theory of intrinsic nonlinear VNE, the next task is to find out how this effect can be probed in experiment. One commonly employed method is to first induce valley polarization in the material, e.g., by optical pumping, and then the VNE will result in a net charge current in the transverse direction, which can be readily detected by a voltage probe~\cite{dau2019Valley}.

Another method, which does not need the creation of valley polarization, is through nonlocal measurement.
Nonlocal transport measurement has been developed as a mature technique for detecting charge-neutral current, such as spin current and valley current~\cite{abanin2009Nonlocal, beconcini2016Nonlocal}. The basic picture is as follows. Taking a sample with a long strip geometry, as shown in Fig.~\ref{fig_nonlocal}.
The valley current in induced locally in some region on the left side by an applied temperature gradient across the strip (along the $y$ direction in Fig.~\ref{fig_nonlocal}). In experiment, this temperature gradient can be exerted by positioning electric heating elements on the sides of the sample, or by shining a laser beam to heat up one side of the strip (see Fig.~\ref{fig_nonlocal}).
Via VNE, the temperature gradient in $y$ direction generates a valley current $j^v_x$ flowing along the strip.
This valley current is detected by a pair of voltage probes at a distance $x$ away from the region where it is generated (see Fig.~\ref{fig_nonlocal}). This detection relies on the inverse valley Hall effect, which converts the valley current along $x$ into a charge current along $y$ in the region between the two probes. The advantages of nonlocal method include: (i) Its setup is relatively simple (without the need of extra device for valley pumping); and (ii) it may help to extract important characteristics of valley transport, such as valley diffusion length.

The theory of nonlocal detection of nonlinear valley current generated from nonlinear valley Hall effect was recently developed in Ref.~\cite{cao2025Theory}. A key discovery there is that the inversion symmetry breaking is essential for the nonlocal measurement. This is because the inverse valley Hall effect (no matter linear or nonlinear), which is required to convert valley current into electric signal, must be odd under inversion~\cite{cao2025Theory}. In such a case,
the linear and nonlinear valley Hall processes must exist simultaneously. To distinguish the different contributions, there are at least three things one can make use of. The first is to modulate the driving with a low frequency $\omega$ (usually less than 100 Hz), and detect the different harmonics in the resulting signal using lock-in amplifier~\cite{kang2019Nonlinear, lai2021Thirdorder}. This can help to separate linear and nonlinear processes into different channels. The second is to vary the distance $x$ between the generation region and detection region, as different contributions may have different decay lengths.
The third is to study the scaling behavior of the resulting signal with respect to local resistivity $\rho=\sigma^{-1}$. Experimentally, $\rho$ is usually varied by changing the temperature $T$. Such analysis has been successfully applied in a recent experiment by He \emph{et al.}~\cite{HeObservation}.

Here, we follow Ref.~\cite{cao2025Theory} to derive the scaling laws for the nonlocal signal due to VNE. The derivation is
very similar to Ref.~\cite{cao2025Theory}, so we will not repeat the details here. As we shall see, although the treatment is similar, the results differ in several important aspects.


As discussed above, we consider a valleytronic system which exhibits both linear VNE, characterized by coefficient $\alpha^v$, and nonlinear VNE, characterized by coefficient $\beta^v$. In the coordinate system of Fig.~\ref{fig_nonlocal}, these coefficients
correspond respectively to the $xy$ and $xyy$ tensor components.

When the driving is modulated with a low frequency $\omega$, one can detect the various harmonics of the nonlocal signal. Here, we are most interested in the second harmonic nonlocal voltage signal $\delta V_{2\omega}$ with frequency $2\omega$, as nonlinear VNE enters this signal.
There are two processes involving VNE that contribute to $\delta V_{2\omega}$: (a) The first is a nonlinear VNE (in generation region) followed by a linear inverse valley Hall process (in detection region); and (b) the second is a linear VNE followed by a nonlinear inverse valley Hall process. Following the analysis in Ref.~\cite{cao2025Theory}, we obtain the following formula for these two contributions:
\begin{eqnarray}
\frac{\delta V^{N}_{2\omega}}{\left(\Delta T\right)^{2}} & = & \frac{e^{-x/\ell_{v}}}{2\pi\ell_{v}}\beta^{v}\sigma^{v}\rho^{2}+\frac{we^{-2x/\ell_{v}}}{8\ell_{v}^{2}}\zeta^{v}\rho^{3}\left(\alpha^{v}\right)^{2},\label{Vnl1}
\end{eqnarray}
where $\Delta T$ is the temperature difference across the strip in the generation region, $w$ is the width of the strip,
$\ell_{v}$ is the valley diffusion length, $\sigma^{v}$ is the linear valley Hall conductivity, and $\zeta^{v}$ is the coefficient for nonlinear inverse valley Hall effect, which actually is the same as the nonlinear charge Hall conductivity~\cite{cao2025Theory}.

Besides $\delta V^{N}_{2\omega}$, there is another contribution. We note that the Nernst experiments are usually performed with open-circuit condition, i.e., in the generation region, the
electric current induced by $\bm\Delta T$ through thermoelectric coefficient $\xi$ cannot flow out of the sample. In this situation, the temperature gradient will also establish an electric field $E=Q\partial_y T$ across the strip in the generation region,
where $Q=\rho\xi$ is the thermopower, also known as Seebeck coefficient. This induced $E$ field will also lead to an additional contribution to nonlocal signal via the electrical valley processes discussed in~\cite{cao2025Theory}. The resulting second harmonic nonlocal signal is given by
\begin{eqnarray}
\frac{\delta V^{Q}_{2\omega}}{\left(\Delta T\right)^{2}} & = & \frac{e^{-x/\ell_{v}}}{2\pi\ell_{v}}Q^2\chi^{v}\sigma^{v}\rho^{2}\nonumber \\
 &  & \qquad+\frac{we^{-2x/\ell_{v}}}{8\ell_{v}^{2}}\zeta^{v}\rho^{3} \left(\sigma^{v}\right)^{2}Q^2.\label{Vnl2}
\end{eqnarray}


Then the total second harmonic signal will be the sum of the two parts
\begin{equation}
  \delta V_{2\omega}=\delta V^{N}_{2\omega}+\delta V^{Q}_{2\omega}.
\end{equation}

To proceed further, we consider the low temperature regime where the linear and nonlinear Mott relations hold, such that
\begin{equation}
  \xi\approx \mathcal L T\sigma',\qquad \alpha^v\approx \mathcal L T(\sigma^v)',\qquad \beta^v\approx \mathcal L T\chi^v.
\end{equation}
Assume that the linear and nonlinear VNEs are dominated by the intrinsic mechanism, so that $\sigma^v,\chi^v\propto \rho^0$.
As for $\zeta^v$, since it is a $\mathcal T$-even process, it does not have an intrinsic contribution. If it is dominated by the Berry curvature dipole mechanism~\cite{Sodemann2015Quantum}, it would scale as $\zeta^v\propto \rho^{-1}$. Substituting these relations into $\delta V_{2\omega}$, we find the following scaling relation
\begin{equation}\label{scaling}
  \frac{\delta V_{2\omega}}{\left(\Delta T\right)^{2}}=\rho^2(c_1+c_2 T^2),
\end{equation}
where $c_1$ and $c_2$ are two temperature-independent scaling parameters. Here, $c_{1}$ captures the contribution from the first term in Eq.~(\ref{Vnl1}), and $c_{2}$ captures the remaining contributions, including Eq.~(\ref{Vnl2}) and the second term of Eq.~(\ref{Vnl1}). In practice, the scaling parameters $c_{i}$ can be extracted by varying the temperature and plotting $\delta V_{2\omega}/\rho^{2}$ against $T^{2}$. The $c_{1}$ and $c_{2}$ can be determined from the intercept and slope of the linear fitting.


If $x\gg l_v\gg w$, the first term in Eq.~(\ref{Vnl1}) [and also Eq.~(\ref{Vnl2})] will dominate over the second term, because of the exponential decay factor. In this case, we have a simple relation
\begin{equation}
    \frac{\delta V^{Q}_{2\omega}}{\delta V^{N}_{2\omega}}\approx \frac{c_2 T^2}{c_1}=\frac{Q^2}{\mathcal L}=\frac{1}{3}(\pi k_BT)^2\left(\frac{\partial \ln\sigma}{\partial \mu}\right)^2.
    \label{ratio}
\end{equation}
As such, the ratio of the two scaling terms in Eq.~(\ref{scaling}) also measures the relative magnitude of nonlinear valley Hall currents induced directly by the temperature gradient and by the Seebeck electric field. Interestingly, this ratio is simply given by the square of thermopower normalized by the Lorenz number. We see that if $Q\ll \sqrt{\mathcal L}=150$ $\mu$V/K, then $\delta V^{Q}_{2\omega}/\delta V^{N}_{2\omega}\ll 1$ and $\delta V_{2\omega}/(\Delta T)^2\rho^2 \approx c_{1}$ should be approximately independent of temperature.
This is likely to be the case at low temperatures. In comparison, if $Q$ is comparable to $\sqrt{\mathcal L}$, then $\delta V^{Q}_{2\omega}$ cannot be ignored, and the $T^2$ term in Eq.~(\ref{scaling}) will be manifested.

The above analysis is for the second harmonic signal $\delta V_{2\omega}$. One can easily see that the dc (rectified) signal $\delta V_0$ should have the same behavior and may also be detected to study the nonlinear VNE.



\section{material example}

\begin{figure}
    \centering
    \includegraphics[width=\linewidth]{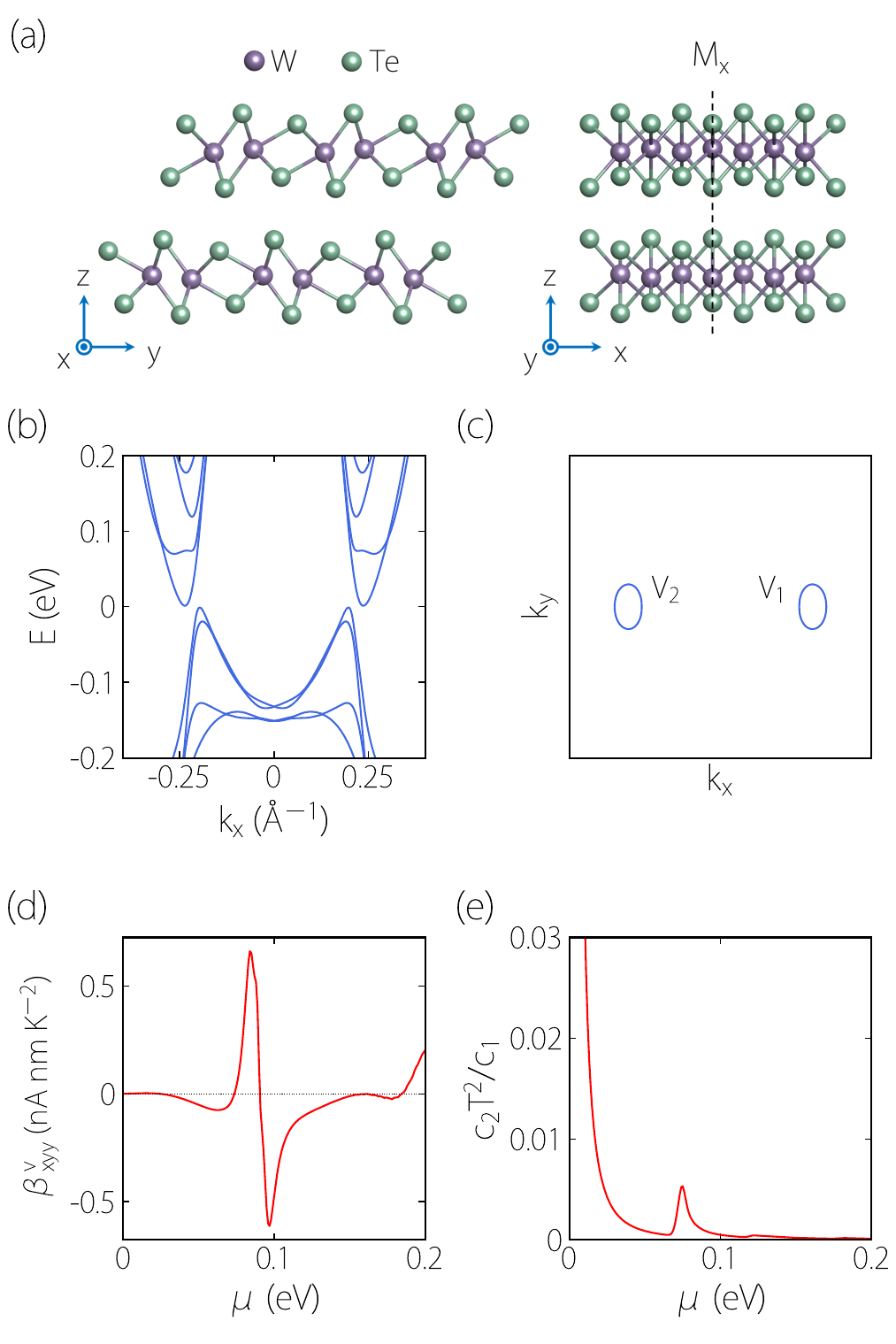}
    \caption{Intrinsic nonlinear VNE in  bilayer WTe$_2$. (a) Crystal structure of bilayer WTe$_2$. It has a mirror $\mathcal M_x$ symmetry, as shown in the right panel. (b) Calculated low-energy band structure, which has two valleys on the $k_x$ axis. (c) Fermi surface at 0.05~eV, which shows the two valleys (labeled as $V_1$ and $V_2$ here). (d) Calculated intrinsic nonlinear VNE conductivity as a function of chemical potential at 10~K. (e) displays the dependence of the ratio $c_2T^2/c_1$ in Eq.~(\ref{ratio}) on the chemical potential.}
    \label{fig_wte2}
\end{figure}

To evaluate the nonlinear VNE in a real material and its manifestation in nonlocal transport, we consider a valley system, the 2D WTe$_{2}$. It has received tremendous interest due to its low structural symmetry. It has been used to explore a range of quantum phenomena, such as the nontrivial topology~\cite{muechler2016Topological}, the quantum spin Hall effect~\cite{zheng2016Quantum,tang2017Quantum,fei2017Edge,wu2018observation,qian2014quantum}, the chirality-dependent photocurrent~\cite{xu2018Electrically}, and the nonlinear Hall effect~\cite{ma2019Observation,kang2019Nonlinear,tiwari2021Giant}.

Here, we consider the bilayer $T_{d}$-WTe$_{2}$ that breaks inversion. The crystal structure of bilayer $T_{d}$-WTe$_{2}$ is shown in Fig.~\ref{fig_wte2}(a).  It has $C_{s}$ point group, which is generated merely by a mirror $\mathcal{M}_{x}$ (see Fig.~\ref{fig_wte2}(a)).
We plot the calculated band structure in Fig.~\ref{fig_wte2}(b). One can see that the system exhibits a semimetal behavior.
Notably, the conduction band exhibits two well separated valleys located on the $\Gamma$-$X$ path, which are connected by $\mathcal{M}_{x}$ symmetry. According to the symmetry analysis summarized in Table~\ref{tab:symmetry}, the nonzero components of the nonlinear VNE conductivity are $\beta_{xyy}^{v}=-\beta_{yxy}^{v}$.

We perform first-principles calculations to evaluate $\beta_{xyy}^{v}$ using Eq.~(\ref{NVNE}) (details in Appendix B). The results are plotted in Fig.~\ref{fig_wte2}(d). One observes that at 10~K, $\beta_{xyy}^{v}$ reaches a peak value of approximately $\mathrm{nA\,nm/K^{2}}$ at $\mu=0.09$~eV. Using this result, we estimate the nonlocal voltage induced by the nonlinear VNE. Previous experimental work reported that $\rho\sim10^{4}\,\Omega$~\citep{ma2019Observation}, and first-principles calculation reported that $\sigma^{v}\sim6\,\mu$S~\citep{cao2025Theory}. We use typical values $x\sim \ell_v \sim 1~\mu$m. Taking a temperature gradient of $40~$K$/\mu$m, the estimated nonlocal voltage is approximately $0.01\,\mathrm{\mu V}$, which is detectable in experiment.

In addition, we evaluate the relative importance of the two terms in Eq.~(\ref{scaling}). 
As shown in Fig.~\ref{fig_wte2}(e), at $\mu=0.09$~eV, where the nonlinear VNE reaches its peak value, the ratio $c_{2}T^{2}/c_{1}\sim 0.001$ is very small. This means that the nonlocal signal is dominated by nonlinear VNE induced directly by the temperature gradient, while the contribution form the Seebeck electric field is negligible.

\section{Discussion and conclusion}

\begin{figure}
    \centering
    \includegraphics[width=0.75\linewidth]{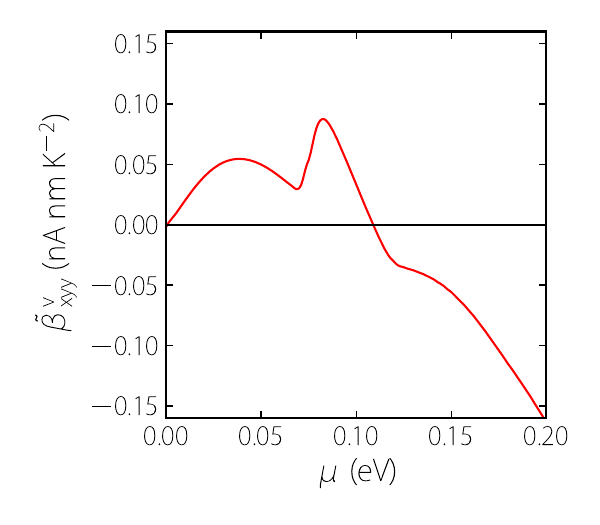}
    \caption{Calculated extrinsic nonlinear VNE conductivity in Eq.~(\ref{extrinsic}) as a function of chemical potential for bilayer WTe$_2$. In the calculation, we take $\tau=10$~fs~\cite{ma2019Observation}.}
    \label{fig_tau2}
\end{figure}

We have focused on the intrinsic nonlinear VNE and discussed its characters and induced signals in nonlocal measurement.
In practice, the intrinsic mechanism may coexist with other extrinsic mechanisms
arising from scattering effects. For example, one such extrinsic
nonlinear VNE was proposed in Ref.~\cite{Yu2014Nonlinear}. It is induced by the off-equilibrium distribution function at the second order of temperature gradient. The corresponding response tensor can be expressed as (derivation in Appendix C)
\begin{equation}
\Tilde{\beta}^v_{abc}=-2\tau^2\int_K\left[d\boldsymbol{k}\right]\frac{v_{a}v_b v_c(\varepsilon_n-\mu)}{T^2}\Big[2f_0'+(\varepsilon_n-\mu)f_0''\Big].
\label{extrinsic}
\end{equation}
This contribution is proportional to the square of scattering time $\tau$, which becomes small when scattering is strong. For example, in bilayer WTe$_2$, $\tau=10$ fs was reported at low temperatures~\cite{ma2019Observation}, and using Eq.~(\ref{extrinsic}), the calculated extrinsic nonlinear VNE is more than one order of magnitude smaller than the intrinsic contribution at $\mu=0.09$ eV (see Fig.~\ref{fig_tau2}). Moreover, even in valley materials with large $\tau$, the contributions from intrinsic and extrinsic nonlinear VNEs can be distinguished by their different scaling behaviors~\cite{Huang2025scaling}. For example, in nonlocal transport, extrinsic VNE $\Tilde{\beta}^v$ exhibits a different scaling relation compared to Eq.~(\ref{scaling}): for $x\gg l_v\gg w$, we get
\begin{equation}
\delta \Tilde{V}_{2\omega}/(\Delta T)^2=\Tilde{c}_{1}+\Tilde{c}_{2}T^{2},
\end{equation}
which does not contain the $\rho^2$ factor.

We have mentioned two methods for detecting nonlinear VNE. One is by valley pumping and electric detection, and the other is by nonlocal measurement. Actually, if the valley carriers possess any valley-contrasted properties, we can utilize such properties for experimental detection. For a valley system with a pair of time reversal connected valleys, if inversion symmetry is broken, the valley carriers will have valley-contrasted orbital magnetic moment $m$~\cite{xiao2007ValleyContrasting}. Then, a valley current flow will lead to opposite magnetization at opposite sample edges, which can be detected, e.g., by magneto-optical methods~\cite{lee2016Electrical}.
On the other hand, if the system preserves inversion symmetry, then $m$ is suppressed, but the carriers may have other valley-contrasted properties. The recently studied anomalous orbital polarizability (AOP) is such an example~\cite{Xiao2021OM,xiao2021adiabatic,Wang2024IPHE}. Under an electric field, carriers with opposite AOP will acquire opposite magnetic moment.

Another perhaps more useful valley-contrasting property in inversion-symmetric valley systems is the magnetic quadrupole moment. It has an orbital part given by~\cite{gao2018Orbital}
\begin{equation}
    q^{\mathrm{O}}_{ab} (\boldsymbol{k}) = \frac23 \text{Re}\sum_{m\neq n} \mathcal{A}_a^{nm} M_b^{mn} - \frac{1}{12} \epsilon_{bcd} \partial_{k_c}\Gamma_{ad},
\end{equation}
where $\boldsymbol{M}^{nm}=\sum_{l\neq m}\frac12 (\boldsymbol{v}^{nl}+\boldsymbol{v}^m\delta_{nl})\times \bm{\mathcal{A}}^{lm}$ is the interband orbital magnetic moment, and $\Gamma_{ab}= \langle u_n | \partial_{k_a}\hat{v}_b | u_n \rangle$ is the Hessian matrix. It also has a spin part given by~\cite{gao2018Microscopic}
\begin{equation}
    q^{\mathrm{S}}_{ab} (\boldsymbol{k}) = \text{Re}\sum_{m\neq n} \mathcal{A}_a^{nm} s_b^{mn},
\end{equation}
where $s_b^{mn}$ is the interband spin magnetic moment. The magnetic quadrupole moment measures the energy variation of an electron induced by a magnetic field gradient. Recent research showed that it is correlated with the nonreciprocal directional dichroism~\cite{Gao2023NDD}. Therefore, when valley carriers are accumulated at sample edges due to the valley current flow from VNE, one can use nonreciprocal directional dichroism to probe this valley polarization in experiment.

In conclusion, we have developed a theory for intrinsic nonlinear VNE. This effect can be generated even in an inversion symmetric system where the linear VNE is prohibited. We show that it is determined by valley BCP dipole, a geometric property of band structure. We further demonstrate a nonlinear Mott relation, which establishes a link between nonlinear VNE and nonlinear valley Hall effect. It states that at low temperatures, the nonlinear VNE is proportional to the nonlinear valley Hall effect, with a constant ratio given by the Lorenz number. Different from the linear Mott relation, the nonlinear relation does not involve a derivative and should have an improved accuracy. We clarify the symmetry characters of nonlinear VNE, and apply our theory to a 2D Dirac model and a concrete material bilayer WTe$_2$. We also propose several methods to detect nonlinear VNE in experiment. Particularly, scaling relations are derived for nonlocal signals induced by VNE, which can help to distinguish different contributions from both linear and nonlinear responses.
These findings lay the foundation for the study of valley caloritronics.

%

\begin{acknowledgements}
  The authors thank D. L. Deng for valuable discussions. We acknowledge support from the Science and Technology Development Fund, Macau SAR (0048/2023/RIB2, 0066/2024/RIA1), the Multi-Year Research Grant, University of Macau (MYRG-GRG2023-00206-IAPME-UMDF), the Start-up Research Grant, University of Macau (SRG2022-00030-IAPME), the Hong Kong Polytechnic University Start-up Fund (P0057929) and Start-up funding from Fudan University.
\end{acknowledgements}

\appendix

\section{Nonlinear Mott Relation}

Here we derive the general nonlinear Mott relation. Since the Fermi–Dirac distribution function $f_0$ depends only on energy, the nonlinear Nernst conductivity can be rewritten from a momentum-space integral to an energy integral. This is achieved by inserting a resolution of identity via the Dirac delta function:
\begin{equation}
f_0'(\varepsilon_n) = \int d\varepsilon \delta(\varepsilon - \varepsilon_n) f_0'(\varepsilon).
\end{equation}
This allows us to decouple the energy-dependent weight from the momentum integral:
\begin{align}
\beta_{abc}(\mu) &= \int [d\boldsymbol{k}] \Lambda_{abc} \frac{(\varepsilon_n - \mu)^2}{T^2} f_0'(\varepsilon_n) \nonumber \\
&= \int [d\boldsymbol{k}] \int d\varepsilon \delta(\varepsilon - \varepsilon_n) \Lambda_{abc} \frac{(\varepsilon - \mu)^2}{T^2} f_0'(\varepsilon). \label{supp_mott}
\end{align}
Here, $\Lambda_{abc} = v_a G_{bc} - v_b G_{ac}$. Exchanging the order of integration yields
\begin{align}
\beta_{abc}(\mu) &= \int d\varepsilon \frac{(\varepsilon - \mu)^2}{T^2} f_0'(\varepsilon) \int [d\boldsymbol{k}] \delta(\varepsilon - \varepsilon_n) \Lambda_{abc} \nonumber \\
&= \int d\varepsilon \frac{(\varepsilon - \mu)^2}{T^2} f_0'(\varepsilon) \chi_{abc}(\varepsilon).
\end{align}
Here, $\chi_{abc}(\varepsilon) = - \int [d\boldsymbol{k}] \delta(\varepsilon - \varepsilon_n) \Lambda_{abc}$ is the zero-temperature nonlinear Hall conductivity at chemical potential $\varepsilon$.

The above expression is the general form of the nonlinear Mott relation, as it establishes the relationship between the intrinsic nonlinear Nernst and Hall conductivity. At temperature close to zero, applying Sommerfeld expansion~\cite{Ziman}, the standard form of nonlinear Mott relation is written as (here $e$ is recovered) :
\begin{equation}
    \beta_{abc}(\mu)=\frac{\pi^{2}k_{B}^{2}}{3e^{2}}\chi_{abc}(\mu).
\end{equation}

\section{Computational Details}

First-principles calculations were performed using the Vienna \textit{ab initio} simulation package (VASP) within the framework of density functional theory (DFT)~\cite{kresse1993Initio, kresse1996Efficiency, kresse1996Efficient}. The projector augmented-wave (PAW) method~\cite{blochl1994Projector} was employed with a plane-wave energy cutoff of 300~eV. Structural optimization was performed within the generalized gradient approximation (GGA) using the Perdew-Burke-Ernzerhof (PBE) functional~\cite{perdew1996Generalized}. To account for van der Waals interactions, the DFT-D3 correction scheme was adopted~\cite{grimme2010Consistent}. The lattice constants are $a = 3.477$~\AA\ and $b = 6.249$~\AA. A vacuum layer of thickness 19~\AA\ was applied. The calculations employed convergence thresholds of $10^{-6}$~eV for total energy and 0.01~eV/\AA\ for atomic forces. The modified Becke–Johnson (mBJ) potential was employed to compute the electronic structure~\cite{Becke2006simple,Tran2009Accurate}, offering improved agreement with experimental band gaps. Spin-orbit coupling (SOC) was included in all calculations. Brillouin zone sampling was performed using a $\Gamma$-centered $12 \times 6 \times 1$ $k$-point mesh. Maximally localized Wannier functions were constructed using the Wannier90 package~\cite{mostofi2008Wannier90}, based on W 5d- and Te 5p-orbital projections. 

\bigskip
\section{$\tau^2$-Contribution to Nonlinear VNE}

Under the relaxation time approximation, the Boltzmann equation for the distribution of electrons in the absence of electric field is given by
\begin{equation}
    f-f_{0}=-\tau\frac{\partial f}{\partial r_{a}}\cdot v_{a},
\end{equation}
where $\tau$ is the relaxation time.

The distribution function up to second order in temperature gradient $\nabla T$ can be written as
\begin{equation}
    f=f_{0}+f_{1}(\partial_{a}T)+f_{2}(\partial_{a}T\partial_{b}T).
\end{equation}
Substituting this expression into the Boltzmann equation, we can obtain the expansion coefficients $f_{n}$ as
\begin{align}
    f_{1}&=-\tau\frac{\partial f_{0}}{\partial r_{a}}\cdot v_{a}, \\
    f_{2}&=-\tau\frac{\partial f_{1}}{\partial r_{b}}\cdot v_{b}.
\end{align}

Rewriting the spatial derivative via the chain rule $\frac{\partial f_{0}}{\partial r_{a}}=\frac{\partial f_{0}}{\partial T}\frac{\partial T}{\partial r_{a}}$, and using the identity $\frac{\partial f_{0}}{\partial T}=-\frac{(\varepsilon-\mu)}{T}f_0'$, one finds that
\begin{align}
f_1 &= \tau v_{a} \frac{\varepsilon_n - \mu}{T} f_0' \partial_a T, \\
f_2 &= \tau^2 \frac{v_{a}v_{b}(\varepsilon_n-\mu)}{T^{2}}\Big((\varepsilon_n-\mu)f_0''+2f_0'\Big)\partial_{a}T\partial_{b}T .
\end{align}

Using this off-equilibrium distribution $f_2$ and evaluating the currents from two valleys, we obtain the $\tau^2$-dependent extrinsic nonlinear valley Nernst conductivity given in Eq.~(\ref{extrinsic}) in the main text.

\bibliographystyle{apsrev4-2}
\bibliography{ref.bib}

\end{document}